\documentclass[floatfix,%
 reprint,showkeys,
 amsmath,amssymb,
 aps,
]{revtex4-2}

\usepackage{graphicx}
\usepackage{bm}
\usepackage{xcolor}
\usepackage[T1]{fontenc}
\usepackage{lmodern}
\usepackage[pdftex, pdftitle={Article}, pdfauthor={Author}]{hyperref}

\usepackage{scalerel}
\usepackage{tikz}
\usetikzlibrary{svg.path}

\definecolor{orcidlogocol}{HTML}{A6CE39}
\tikzset{
  orcidlogo/.pic={
    \fill[orcidlogocol] svg{M256,128c0,70.7-57.3,128-128,128C57.3,256,0,198.7,0,128C0,57.3,57.3,0,128,0C198.7,0,256,57.3,256,128z};
    \fill[white] svg{M86.3,186.2H70.9V79.1h15.4v48.4V186.2z}
                 svg{M108.9,79.1h41.6c39.6,0,57,28.3,57,53.6c0,27.5-21.5,53.6-56.8,53.6h-41.8V79.1z M124.3,172.4h24.5c34.9,0,42.9-26.5,42.9-39.7c0-21.5-13.7-39.7-43.7-39.7h-23.7V172.4z}
                 svg{M88.7,56.8c0,5.5-4.5,10.1-10.1,10.1c-5.6,0-10.1-4.6-10.1-10.1c0-5.6,4.5-10.1,10.1-10.1C84.2,46.7,88.7,51.3,88.7,56.8z};
  }
}

\newcommand\orcidicon[1]{\href{https://orcid.org/#1}{\mbox{\scalerel*{
\begin{tikzpicture}[yscale=-1,transform shape]
\pic{orcidlogo};
\end{tikzpicture}
}{|}}}}

\begin{document}

\title{Beyond Comoving Volume: Horizon Flux and Matter Creation in Modified Cosmology from the Unified First Law of Thermodynamics
}

\author{V\'{\i}ctor H. C\'ardenas\orcidicon{0000-0002-9788-3967}}
\email{victor.cardenas@uv.cl} 
\affiliation{Instituto de F\'{\i}sica y Astronom\'{\i}a, Facultad de Ciencias, Universidad de Valpara\'{\i}so Av. Gran Breta\~na 1111, Valpara\'{\i}so, Chile}

\author{Miguel Cruz\orcidicon{0000-0003-3826-1321}}
\email{miguelcruz02@uv.mx}
\affiliation{Facultad de F\'{\i}sica, Universidad Veracruzana 91097, Xalapa, Veracruz, M\'exico}

\author{Samuel Lepe\orcidicon{0000-0002-3464-8337}}
\email{samuel.lepe@pucv.cl}
\affiliation{Instituto de F\'\i sica, Pontificia Universidad Cat\'olica de Valpara\'\i so, Casilla 4950, Valpara\'\i so, Chile}%

\begin{abstract}
We explore the derivation of the Friedmann equations from a thermodynamic perspective, applying the unified first law of thermodynamics to the apparent horizon of a flat Friedmann-Lemaître-Robertson-Walker (FLRW) universe. We extend this framework to incorporate gravitationally induced particle creation, treating the region enclosed by the apparent horizon as an open thermodynamic system. A crucial aspect of our analysis is the recognition that the apparent horizon volume is not comoving; this requires consistent accounting of particle exchange across the moving boundary. We demonstrate that the evolution of the particle number, and explicitly the matter entropy, can be decomposed into two distinct physical contributions: genuine bulk particle production and a net flux induced by the dynamics of the horizon itself. Finally, we derive the Generalized Second Law (GSL) in this setting, showing transparently how the total entropy budget is balanced by horizon thermodynamics, bulk creation, and boundary fluxes.
\end{abstract}

\date{\today}

\keywords{matter creation, entropic cosmology, apparent horizon}
\maketitle


\section{\label{sec:level1} Introduction}

The standard cosmological model, $\Lambda$CDM, has provided a remarkable description of the large scale structure and expansion history of the universe \cite{scott2018, peebles}. However, it depends on a cosmological constant $\Lambda$ whose basic physical origin remains obscure, because there is an enormous difference between the measured value and the one predicted from quantum theory \cite{Carroll_2001}. Furthermore, the model currently faces significant challenges, most notably the Hubble tension, a statistically significant discrepancy between early universe estimations, such as those from the Planck satellite, and late universe local measurements using Cepheid variables and Type Ia supernovae, to mention some \cite{planck, Riess_2022}. These limitations have motivated the exploration of alternative mechanisms that can account for late time acceleration while simultaneously addressing these observational discrepancies \cite{Valentino_2021}.

Among these alternatives, gravitationally induced particle creation has emerged as a compelling scenario. A seminal work in this context can be found in \cite{PhysRevLett.21.562}, which demonstrated that elementary particles can be produced in an expanding universe even in the absence of an initial particle density. Comparable concepts are discussed in reference \cite{BROUT1978}. A detailed formulation of the mechanism within the framework of non-equilibrium thermodynamics was provided in \cite{1989}; see also \cite{lima2007}; in this scenario, the gravitational field is assumed to transfer energy into the matter sector, producing a negative creation pressure that effectively reproduces the behavior of dark energy with no need for extra contributions coming from an exotic type of matter \cite{PhysRevD.66.103502, Ramos_2014}. The phenomenological viability of such models, including their capacity to accurately reproduce specific features of the expansion history like the transition between decelerated and accelerated phases, has been widely investigated \cite{chakraborty2014third, chakraborty2014cosmological}. Specifically, a recent work has established that this creation pressure, arising from an adiabatic transfer of energy from gravitation to matter in an open thermodynamic system, can successfully emulate the background dynamics of various dark energy models, including $\Lambda$CDM and $w$CDM, using only a single pressureless fluid component \cite{Cardenas_2024}. Although these models work well at the phenomenological level, they still need a solid thermodynamic basis to guaranty compatibility with the principles of general relativity; see, for instance, Ref. \cite{gsl}, where the thermodynamic formalism for this type of cosmological model has been investigated in the late time regime. Moreover, the viability of matter creation models has been recently subjected of observational scrutiny. Current literature extensively constrains these scenarios using robust statistical methodologies. These studies highlight the necessity of establishing clear functional forms for the creation rate to perform goodness-of-fit evaluations against the standard cosmological model \cite{Gohar_2026, Bhattacharjee_2026, Halder_2025, Gohar_2021}.

A main result of the implementation of the aforementioned scenario in theoretical cosmology is the thermodynamic derivation of the Einstein field equations. Pioneered by Jacobson \cite{jacobson1995thermodynamics} and extended to cosmological settings in \cite{Hayward_1998}, this approach interprets the Friedmann equations not as imposed dynamical laws but as thermodynamic equations of state arising from the application of the unified first law of thermodynamics, $TdS = -dE + WdV$, to the apparent horizon of the universe \cite{Cai:2005ra,Akbar:2006kj,Cai:2006rs}.

In this work, we connect these two frameworks, entropic cosmology and matter creation, by treating the region enclosed by the apparent horizon as an open thermodynamic system. A crucial but often overlooked subtlety in this approach is the definition of the volume. Unlike standard fluid dynamics, which tracks comoving volumes, the apparent horizon is a dynamical boundary that is \textit{not} comoving. Consequently, there is an inherent particle flux across the horizon whenever the universe decelerates or accelerates. Then, this ``horizon flux'' is not just a minor geometric feature of spacetime, but it plays a crucial role in the thermodynamic description of the universe. By deriving the GSL for this non-comoving volume, we find that there is a crucial relationship between the particle production rate and the Hubble parameter in order to maintain thermodynamic consistency while at the same time allowing for a phase of accelerated expansion in the universe. This has immediate consequences for the viability of matter creation models.

According to recent observational bounds from \cite{Schiavone2026}, particle creation models align with data only when the produced component has an equation of state $w<-1/3$, ruling out pressureless cold dark matter ($w=0$) as an explanation for cosmic acceleration. Crucially, however, such analysis neglects the contribution of the apparent horizon. By incorporating this thermodynamic element, our formulation adheres to the fundamental objective of the matter creation framework: to explain the observed cosmic acceleration without relying on exotic matter contributions. As our analysis will demonstrate, pressureless cold dark matter can consistently account for an accelerated expansion of the universe at late times.

The work is organized as follows: In Section \ref{sec: level2}, we review the entropic derivation of the Friedmann equations in a flat FLRW spacetime, and in Section \ref{sec:matt} we address how matter content can be incorporated into this framework. Section \ref{sec:creation} is devoted to examining the role of matter creation within this setting, where we determine the corresponding thermodynamic description of the model using horizon-flux-corrected expressions. Finally, in Section \ref{sec:final}, we present the conclusions of the study. Appendix \ref{sec:app} is dedicated to assessing the physical viability of our framework. By employing the cosmic chronometers dataset, we reconstruct some key model quantities within the Gaussian Processes approach. Throughout this work, we adopt natural units where $G=c=M_{\rm{pl}}=1$ and the following sign convention $(- + + \,+)$.

\section{\label{sec: level2} modified cosmology from the unified first law of thermodynamics}
In this section, we revisit the derivation of the Friedmann equations from a thermodynamic perspective by applying the unified first law of thermodynamics, as formulated by Hayward, to the apparent horizon of a flat FLRW universe. Following the approach of Akbar and Cai, we define the energy $E$ as the Misner-Sharp mass enclosed within the apparent horizon \cite{Cai:2005ra,Akbar:2006kj,Cai:2006rs}. The goal is to show the equivalence between the unified first law of thermodynamics, when applied to the apparent horizon of an FLRW universe, and the Einstein field equations. This approach is based on the assumption that the apparent horizon has an entropy proportional to its area and a temperature determined by its surface gravity.
We begin with the unified first law of thermodynamics applied at the apparent horizon \cite{quevedo, faraoni}
\begin{equation} \label{eq: dE}
T dS_h = -dE + W dV,
\end{equation}
where $E=\rho V$ is the total energy inside the horizon, $W = (\rho - p)/2$ is the work density, $V = 4\pi r_A^3/3$ is the volume enclosed by the apparent horizon $r_A=H^{-1}$ in a flat universe, $S_h$ is the entropy associated with the horizon, and $T = \kappa/(2\pi)$ is the temperature of the horizon, where $\kappa$ is the standard Kodama-Hayward surface gravity \cite{kodama1980conserved}, given by
\begin{equation}\label{eq: kappa}
    \kappa = -\frac{1}{r_A}\left( 1 - \frac{\dot{r}_A}{2Hr_A} \right).
\end{equation}
Some remarks concerning the temperature are now appropriate. The surface gravity can be written in the following way in terms of the cosmographic parameter $q:=-1-\dot{H}/H^{2}$ \cite{cruzlepe}
\begin{equation}
    \kappa = -\frac{1}{2r_{A}}(1-q).
\end{equation}
This latter expression shows that, in an expanding Universe, we obtain $\kappa < 0$ because $q<0$. Strictly speaking, the sign of the temperature associated with the apparent horizon is determined by the specific type of horizon. As explained in Refs. \cite{helou1,helou2,posit}, an expanding cosmology corresponds to a past-inner trapping horizon. In this situation, we must impose $T \propto -\kappa$ together with $\kappa < 0$, as noted above for the surface gravity. Under these conditions, the resulting physical temperature is positive and is given by
\begin{equation}
    T = -\frac{\kappa}{2\pi} = \frac{1}{2\pi r_{A}}\left(1-\frac{\dot{r}_{A}}{2Hr_{A}}\right) = \frac{1}{4\pi r_{A}}(1-q). \label{eq:temp}
\end{equation}
Notice that the horizon temperature given by $T=H/2\pi$ corresponds to the case $q=-1$. The expression (\ref{eq: dE}) can be interpreted as a balance law for energy changes inside the horizon, accounting for pressure work and heat exchange with the horizon. The expansion of the universe sets a characteristic time scale, with the expansion time defined as $H^{-1} \equiv a/\dot{a}$. A given particle species will remain in thermal equilibrium with the cosmic fluid as long as its interaction rate is sufficiently high to keep up with the decreasing temperature. This is physically justified provided the thermodynamic relaxation time-scale of the cosmic fluid, $\tau$, is significantly smaller than the Hubble time-scale, $\tau \ll H^{-1}$. In the late universe, the microscopic interaction rates of the fluid components vastly exceed the expansion rate $H$, ensuring that local thermodynamic equilibrium is maintained and allowing the bulk to remain dynamically coupled to the horizon \cite{maartens}. However, for a weakly interacting component such as dark matter, it is not enough to invoke rapid microscopic processes to guarantee local thermodynamic equilibrium. On cosmological scales, collisionless dark matter can instead be modeled dynamically as an effective perfect fluid, whose behavior is governed by collective gravitational effects. In this {\it coarse-grained} regime, macroscopic thermodynamic quantities remain well-defined and stable, enabling the bulk fluid to be coupled to the cosmic expansion without relying on assumptions of high microscopic scattering rates \cite{Sussman_2016}.

Taking the right hand side of (\ref{eq: dE}) as a time derivative and using the conservation equation
\begin{equation}\label{eq: cons}
    \dot{\rho} + 3H(\rho + p) = 0,
\end{equation}
we get 
\begin{equation}
    TdS_h = \frac{8 \pi^2 (\rho + p) }{H^3} \left(\frac{H}{2\pi} \left[ 1 + \frac{\dot{H}}{2H^2}\right] \right)dt,
\end{equation}
which can be identified with the temperature (\ref{eq:temp}) when $r_A = H^{-1}$, while the remaining part should correspond to the time derivative of the entropy
\begin{equation}
    \dot{S}_h = \frac{8 \pi^2 (\rho + p) }{H^3}.
\end{equation}
To close the equivalence, we use the Bekenstein-Hawking entropy
\begin{equation}
    S_h = \frac{4\pi r_A^2}{4} = \frac{\pi}{H^2} \rightarrow \dot{S}_h = -\frac{2\pi}{H^3}\dot{H},\label{eq:bh}
\end{equation}
solving this by equating both time derivatives of the horizon entropy leads to the second Friedmann equation
\begin{equation}
\dot{H} = -4\pi (\rho + p)= \frac{4\pi}{3H}\dot{\rho},\label{eq:acc}
\end{equation}
where the conservation equation (\ref{eq: cons}) was used in the second equality. By integrating this result one recovers the standard Friedmann equation:
\begin{equation}
H^2 = \frac{8\pi }{3} \rho + \text{const.}
\end{equation}
In this setup, the resulting integration constant can effectively act as a cosmological constant, thereby addressing the fundamental issue of vacuum energy. It is important to emphasize that the identification $E=\rho V$ is not postulated by convenience. Instead, comes from the definition of the Misner-Sharp mass, $E=r_{A}/2=1/(2H)$. The application of the unified first law yields the acceleration equation (\ref{eq:acc}) which in turn results in the Friedman equation after integration. It is only after multiplying this derived density, $\rho = 3H^{2}/(8\pi)$, by the geometric volume, $V = 4\pi/(3H^{3})$, that the relation $\rho V = 1/(2H)$ follows in a natural way. Then, the equivalence $E=\rho V$ is a thermodynamic consequence for any cosmological model. A subtle but important point in this derivation is the choice of volume. In standard cosmology, the conservation equation (\ref{eq: cons}) follows from energy-momentum conservation and is usually expressed using a comoving volume scaling as $V \propto a^{3}$. In that situation, the relation can be written as $d(\rho a^{3}) + p\,d(a^{3}) = 0$, which describes the adiabatic expansion of the cosmic fluid. However, in a horizon-thermodynamics framework, it is more natural to use the volume bounded by the apparent horizon $V = \tfrac{4\pi}{3} r_{A}^{3}$, because it identifies the portion of the universe that is causally accessible to the observer.

The key point is that fluid variables such as $\rho$, $p$, and $n$ are defined as densities within a physical volume. Meanwhile, entropy $S_h$ is fundamentally associated with the area of the boundary, which is determined by the apparent horizon. This justifies writing the total energy as $\rho V$, where $V$ is the volume enclosed by the horizon, ensuring consistent thermodynamic treatment.

In later sections, we will show that using this same volume in the matter sector allows us to derive a modified conservation law and subsequently incorporate matter creation in a consistent way.

\subsection{Introducing matter}
\label{sec:matt}

To examine the GSL of thermodynamics, the thermodynamics of the apparent horizon must be supplemented by the contribution of the matter sector. In this section, we build on the previous discussion by incorporating the thermodynamics of the cosmic fluid, now allowing for a variable particle number. In this situation, the unified first law reads \cite{callen}
\begin{equation}\label{eq: tdsm}
    T_m\, dS_m = d(\rho V) + p\, dV - \mu\, d(nV),
\end{equation}
where $n$ denotes the number density, $N=nV$ the total number of particles, and $\mu$ characterizes the chemical potential. For consistency with the horizon-based framework adopted above, we identify $V$ with the volume enclosed by the apparent horizon of radius $r_A$. As argued in \cite{Nojiri:2024zdu, PhysRevD.109.103515}, the variation in particle number linked to horizon crossing can be accounted for within the matter–entropy balance. Within this prescription the total entropy of the observable universe, delimited by the apparent horizon, is given by the sum of the entropy of the horizon and that of the matter fields contained within it, $S_{\mathrm{tot}}= S_{h}+S_{m}$. Consequently, the GSL can be written as
\begin{equation}
T\dot{S}_h + T_{m}\dot{S}_m \geq 0,
\end{equation}
holds as long as the microscopic dynamics of the matter fields obey conventional thermodynamic principles and $T$ and $T_{m}$ represent the temperature of the apparent horizon and matter fields, respectively. Following the findings reported in \cite{Mimoso_2016}, we impose the condition $T = T_{m}$, which holds for non-relativistic matter as well as for dark energy. In what follows, we make explicit the assumptions underlying the use of (\ref{eq: tdsm}) and the associated definition of the chemical potential. 
Our aim is not to obtain a new microscopic formula for $\mu$, but rather to guaranty the thermodynamic consistency of the matter sector when the \emph{system} is taken to be the region bounded by the apparent horizon, $V_h=\frac{4\pi}{3}r_A^{3}$. Because $r_A(t)$ is time dependent, the corresponding volume is not comoving, and the total number of particles inside it,
\begin{equation}
N_{h}(t)=n(t)\,V_h(t), \label{eq:total}
\end{equation}
can change in a generic way even when no particles are being produced. Throughout this subsection, we therefore adopt the \emph{no-creation} hypothesis: the microscopic dynamics does not generate particles in the bulk, and the cosmic fluid remains in local equilibrium. In particular, we assume that the matter sector evolves adiabatically,
\begin{equation}
dS_m=0,
\end{equation}
so that any change in internal energy and mechanical work associated with the evolution of $V_h$ is balanced by the exchange of particles across the moving boundary, encoded in the chemical potential term \footnote{Equivalently, the matter sector is taken to have vanishing entropy production; possible non-equilibrium contributions associated with the horizon can be absorbed into the horizon entropy $S_h$ and will be constrained by the GSL.}.

Under these assumptions, the Gibbs relation for an open system,
\begin{equation}
T_m\, dS_m = d(\rho V_h) + p\, dV_h - \mu\, d(nV_h),
\label{eq: GibbsOpen}
\end{equation}
reduces to
\begin{equation}
V_h \dot{\rho} + (\rho+p)\dot V_h = \mu\,\dot N.
\label{eq: adiabatic_open}
\end{equation}
Using $\dot N=\dot n\,V_h+n\,\dot V_h$ and $\dot V_h/V_h=3\,\dot r_A/r_A$, Eq.~(\ref{eq: adiabatic_open}) can be written in compact form
\begin{equation}
\dot{\rho} + (\rho+p)\frac{\dot V_h}{V_h} = n\mu\left(\frac{\dot n}{n} + \frac{\dot V_h}{V_h}\right).
\label{eq: mu_general_Vh}
\end{equation}
Notice that $3\,\dot r_A/r_A=3H(1+q)$ where $1+q= 
-\dot{H}/H^{2}$ is the deceleration parameter. To make contact with the conventional cosmological limit, observe that when the system is chosen as a comoving volume $V \propto a^{3}$ and the matter sector is closed ($\dot N = 0$), Eq.~(\ref{eq: GibbsOpen}) with $dS_m = 0$ simplifies to the standard adiabatic conservation law,
\begin{equation}
d(\rho a^{3})+p\,d(a^{3})=0 \quad \Longleftrightarrow \quad \dot\rho=-3H(\rho+p).
\end{equation}
In the current horizon-volume framework, by contrast, $\dot N$ is not required to be zero because the boundary is time-dependent; the associated exchange of particles is exactly what the $\mu\,dN$ term represents.

Finally, in local equilibrium, the chemical potential is fixed by the Gibbs--Duhem relation. For a fluid with negligible entropy density (or, equivalently, $Ts\ll \rho$), one obtains the standard expression
\begin{equation}
\mu=\frac{\rho+p}{n},
\label{eq: mu_enthalpy}
\end{equation}
i.e.\ the enthalpy per particle. As a consistency check, in the dust limit $p\simeq 0$ and $\rho\simeq m n$, Eq.~(\ref{eq: mu_enthalpy}) implies $\mu\simeq m$, as expected for a cold non-relativistic matter component. Equation (\ref{eq: mu_enthalpy}) will be used in the following as the equilibrium closure relating $\mu$ to the macroscopic variables $(\rho,p,n)$ within the horizon-volume thermodynamic framework. When there is no particle production, that is, $n$ obeys the usual conservation equation $\dot{n} + 3Hn = 0$, the standard energy conservation law (\ref{eq: cons}) is restored.
This formulation sets the stage for the introduction of irreversible processes in the next section, where we allow for gravitationally induced particle creation.

\section{Matter creation}
\label{sec:creation}

Whereas the previous section formulated an idealized setup based on adiabatic local equilibrium ($dS_m=0$), gravitationally induced particle creation violates this condition in an essential way. The region enclosed by the apparent horizon must now be treated as an open thermodynamic system undergoing irreversible processes; we must therefore adopt a generalized Clausius relation that incorporates internal entropy production, $d_{i}S>0$ \cite{1989}. Once particle production is allowed, the number of particles is no longer conserved, even within a comoving volume. The fluid must therefore be characterized using the formalism of irreversible thermodynamics \cite{eling2006non, maartens}. Locally, we consider that macroscopic dynamics allow deviations from the local equilibrium pressure as a result of the contribution from the {\it creation pressure} $p_{c}$, then
\begin{equation}
\dot{\rho}+3H(\rho+P)=0,
\label{eq:rho_creation}
\end{equation}
where the total pressure of the fluid is decomposed as $P=p+p_{c}$. Additionally, the balance equation for the particle number density takes the form,
\begin{equation}
\dot{n}+3Hn=\Gamma n,
\label{eq:n_creation}
\end{equation}
which emerges from the consideration of the non-conservation for the vector particle number $N^{\mu}_{;\mu}=n\Gamma$ with $N^{\mu}=nu^{\mu}$ in a FLRW universe. $\Gamma$ denotes the particle production rate, with $\Gamma > 0$ ensuring that particles are indeed generated, and $p_c = p_c(\Gamma)$ representing the effective pressure term that incorporates the backreaction due to this creation. Consequently, the work density evaluated at the apparent horizon is modified to include this effective pressure, for details see \footnote{It is widely recognized that the flat FLRW metric can also be written in an equivalent form using a warped product as
\begin{equation}
    ds^{2} = h_{ab}dx^{a}dx^{b} + R^{2}(t)(d\theta^{2}+\sin^{2}\theta d\varphi^{2}),
\end{equation}
where $R(t,r)$ denotes the physical radius of the FLRW universe, given by $R(t,r):=a(t)r$, with $r$ the comoving coordinate and $a(t)$ the scale factor. Furthermore, we introduce the metric $h_{ab} := \mbox{diag}[-1,a^{2}(t)]$, where in this context $a,b=0,1=t,r$. If the matter content of the universe is modeled by a perfect fluid with energy-momentum tensor $T_{ab} = (\rho + P)u_a u_b + P g_{ab}$, then the work density $W$ associated with these matter fields is defined in terms of the energy density and pressure as
\begin{equation}
    W = -\frac{1}{2} T = \frac{1}{2}(\rho - P),\label{eq:work}
\end{equation}
where $T$ denotes the two-dimensional normal trace of the energy-momentum tensor, $T := h^{ab} T_{ab}$.}:
\begin{equation}
W_{\text{eff}} = \frac{1}{2}(\rho - p - p_c),
\end{equation}
substituting this generalized work density into the unified first law applied to the apparent horizon, we obtain:
\begin{equation}
T dS_h = -d(\rho V_{h}) + \frac{1}{2}(\rho - p - p_c)dV_{h}.
\end{equation}
Notice that the creation pressure naturally modifies the work density at the apparent horizon. Instead of representing a pure thermodynamic heat flow, the additional term $-(1/2)p_{c}dV_{h}$ accounts for the effective mechanical work exerted on the boundary by the irreversible bulk particle creation. Taking the time derivative of this generalized relation and using the modified energy conservation equation (\ref{eq:rho_creation}) together with the temperature for the horizon discussed before, the evolution of the horizon entropy becomes explicitly dependent on the creation pressure
\begin{equation}
\dot{S}_h = \frac{8\pi^2}{H^3} (\rho + p + p_c).
\end{equation}
If we compare this expression to the time derivative of the Bekenstein-Hawking entropy, as before, we successfully recover the modified acceleration equation \cite{lima2007, PhysRevD.66.103502, Ramos_2014, Cardenas_2024}
\begin{equation}
    \dot{H} = -4\pi (\rho + p + p_{c}),\label{eq:pc}
\end{equation}
confirming that modified Friedmann dynamics emerges from the generalized description of the apparent horizon; to obtain this last expression, we have used the equation (\ref{eq:bh}). Correspondingly, the matter sector within the apparent horizon can no longer be described by the equilibrium Gibbs relation. Incorporating Prigogine's formulation for open cosmological systems, the total variation of the matter entropy, $dS_m$, must be decomposed into a reversible entropy exchange flux across the apparent horizon, $d_e S$, and an internal entropy production strictly due to genuine particle creation, $d_i S \ge 0$. The generalized Gibbs equation becomes
\begin{equation}
T_m dS_m = d(\rho V_{h}) + p dV_{h} - \mu d(n V_{h}) + T_m d_i S.
\end{equation}
Here, the chemical potential $\mu$ determines how changes in particle number are thermodynamically linked to entropy production. In a conventional open system, the term $\mu d(n V)$ represents the reversible energy exchange associated with particles crossing a boundary. In our framework, however, the total change in particle number accounts for both the reversible geometric flux across the evolving horizon and the irreversible particle production process encoded in $\Gamma$.
The internal entropy production is driven exclusively by the latter. The amount of irreversible heat produced is directly proportional to both the thermodynamic cost of particle production, $\mu$, and the rate of particle creation:
\begin{equation}
T_m d_i S = \mu (\Gamma n V_{h}) dt.
\end{equation}
According to the second law, variations in the number of particles satisfying $dN=d(nV_{h}) > 0$ are allowed. This implies that, within the framework of open systems and irreversible processes, spacetime is capable of generating matter, while the reverse process is not allowed. A useful definition is given by the entropy per particle $\sigma$, with $s = n\sigma$ being the entropy density. Under the adiabatic creation condition ($\dot{\sigma} = 0$), newly produced particles are generated in perfect thermal equilibrium with the cosmic fluid. Accordingly, as we will demonstrate below, the entropy per particle remains constant, and all internal entropy production arises from the irreversible increase in the number of particles in the bulk, thereby ensuring $\dot{S}_m > 0$ without violating local thermodynamic equilibrium.

As discussed above, according to the horizon-thermodynamic description, the relevant system is the region enclosed by the apparent horizon given by the volume $V_h=(4\pi/3)r_A^3$, and the total number of particles in this region is given by equation (\ref{eq:total}).
Differentiating and using (\ref{eq:n_creation}) yields a clean split into two physically distinct contributions,
\begin{equation}
\dot N_h
=\underbrace{\Gamma N_h}_{\text{bulk creation}}
+\underbrace{ 3HqN_h}_{\text{horizon flux}},
\label{eq:Nh_split}
\end{equation}
where $3Hq=\dot{V}_h/V_h - 3H$. The first term is genuine particle production within the fluid, while the second term accounts for the fact that $V_h(t)$ is not comoving: even if $\Gamma=0$, $N_h$ changes whenever $\dot V_h\neq 3H V_h$, which can be interpreted as a net particle flux across the moving horizon. Alternatively, Eq. (\ref{eq:Nh_split}) shows that the fractional growth rate of the particle number $\dot{N}_{h}/N_{h}$, is simply the sum of the creation rate $\Gamma$ and the horizon slip rate $3Hq$.

Consequently, the matter entropy contained within the horizon is
\begin{equation}
S_m(t)=s(t)\,V_h(t)=\sigma(t)\,N_h(t).
\end{equation}
Therefore, the time variation of $S_m$ can always be decomposed as
\begin{equation}
\dot S_m=\underbrace{\sigma\,\dot N_h}_{\text{changing }N_h \ \text{contribution}}
+\underbrace{N_h\,\dot\sigma}_{\text{non-adiabaticity per particle}}.
\label{eq:Sm_general_split}
\end{equation}
For open thermodynamic systems, beginning with the basic definition of an adiabatic process, $dQ = 0$, and applying it to the Gibbs relation, one can explicitly show that the specific entropy $\sigma$ must remain constant \cite{balfagon2015accelerated}. In addition, the second law implies that such a process allows only particle production and strictly rules out particle annihilation. Consequently, the condition of \emph{adiabatic creation}, $\dot{\sigma}=0$, is not an arbitrary assumption but follows directly from these fundamental principles, guaranteeing local equilibrium and zero entropy production per particle \cite{BARROW1988743, Zimdahl_1993, CALVAO1992223}; hence, $\dot{S}_{m} \neq 0$. In that case, from Eq. (\ref{eq:Nh_split}) we have
\begin{equation}
\dot S_m=\sigma\,\dot N_h
= \sigma N_h \left( \Gamma + 3Hq \right),
\label{eq:Sm_adiabatic_creation}
\end{equation}
clearly defining that the matter entropy inside the apparent horizon changes through:
\begin{itemize}
    \item[(I)] genuine bulk creation and
    \item[(II)] net flux induced by the moving boundary.  
\end{itemize}
It is worth mentioning that contribution (II) corresponds to the outward (inward) flux across the horizon identified in \cite{PhysRevD.109.103515} for matter fields contained within it during periods of accelerated (decelerated) expansion. On the other hand, it is important to note that in the decelerating regime ($q>0$), both particle production in the bulk and horizon flux contribute positively to the increase in matter entropy. Once the universe transitions to the accelerating regime $(q<0)$, the horizon flux term becomes negative. In physical terms, this means that the expansion of the apparent horizon slows down relative to cosmic fluid flow. Consequently, for $\dot{S}_{m}\geq 0$ to hold at late times, where $q<0$, the bulk creation rate $\Gamma$ has to satisfy the condition
\begin{equation}
    \Gamma \geq-3Hq.\label{eq:decelcond}
\end{equation}
To rigorously establish the magnitude of this bound during the late-time acceleration regime ($q<0$), we can connect this thermodynamic requirement directly with the macroscopic kinematic background. As demonstrated in Eq. (\ref{eq:q}), the modified Friedmann dynamics for a pressureless fluid ($p=0$) dictate that $q = (1/2)(1-\Gamma/H)$, which yields the exact dynamical requirement $\Gamma = H(1-2q)$. For this dynamically required rate to satisfy the GSL condition $\Gamma \geq -3Hq$, the deceleration parameter is strictly constrained to $q \geq -1$. By evaluating the saturation limits of this relation we observe that at the onset of acceleration ($q \to 0$), the rate approaches $\Gamma \to H$, whereas at the asymptotic de Sitter limit ($q \to -1$), it saturates exactly at $\Gamma \to 3H$. Consequently, for any viable accelerating phase ($-1 \leq q < 0$), the kinematic requirements naturally ensure that the production rate is bounded by $H < \Gamma \leq 3H$. This proves that $\Gamma > H$ is a robust, conservative lower bound demanded simultaneously by background dynamics and thermodynamic consistency. A convenient way to connect $\Gamma$ and $p_c$ is obtained from the local Gibbs relation (per unit volume),
\begin{equation}
nT_m\,\dot\sigma=\dot\rho-\frac{\rho+p}{n}\,\dot n,
\label{eq:gibbs_local}
\end{equation}
which, upon using (\ref{eq:rho_creation})--(\ref{eq:n_creation}), gives
\begin{equation}
nT_m\,\dot\sigma=-3H\,p_c-\Gamma(\rho+p).
\label{eq:sigma_dot_creation}
\end{equation}
Thus, adiabatic creation ($\dot\sigma=0$) implies the standard relation
\begin{equation}
p_c=-\frac{\rho+p}{3H}\,\Gamma,
\label{eq:pc_adiabatic}
\end{equation}
demonstrating that the effective creation pressure is precisely what enforces $\dot\sigma=0$ at the macroscopic level. The generalized entropy is
\begin{equation}
S_{\rm tot}=S_h+S_m,
\qquad\Rightarrow\qquad
\dot S_{\rm tot}=T\dot S_h+T_{m}\dot S_m.
\end{equation}
The horizon contribution $\dot S_h$ is obtained from the horizon unified first-law/Clausius relation used in the previous section, with the replacement
\begin{equation}
\rho+p\;\longrightarrow\;\rho+p+p_c,
\end{equation}
since particle creation modifies the effective pressure appearing in the energy flux through the horizon. 
The matter contribution $\dot S_m$ is given by (\ref{eq:Sm_general_split}), or by (\ref{eq:Sm_adiabatic_creation}) in the adiabatic creation case.

Altogether, the GSL can be written in a form that makes the two sources of matter-entropy change explicit,
\begin{equation}
T \left\lbrace \dot S_h
+\sigma\left[\Gamma N_h+n\!\left(\dot V_h-3H V_h\right)\right]
+N_h\,\dot\sigma \right\rbrace
\;\geq\;0,
\label{eq:GSL_with_creation}
\end{equation}
where we have considered $T=T_{m}$ and $\dot\sigma$ can be removed in favour of $(p_c,\Gamma)$ by using (\ref{eq:sigma_dot_creation}). Equation (\ref{eq:GSL_with_creation}) makes it clear that the generalized entropy production receives contributions from:
\begin{itemize}
    \item[(I)] the horizon sector ($S_h$),
    \item[(II)] bulk particle creation ($\propto \Gamma N_h$),
    \item[(III)] particle/entropy flux associated with the dynamical horizon volume ($\propto \dot V_h-3HV_h$), and
    \item[(IV)] possible non-adiabaticity per particle ($\dot\sigma\neq 0$).
\end{itemize}
Explicitly,
\begin{equation}
\dot{S}_{\rm tot}=\frac{1}{2}(1-q^{2})+\frac{H(1-q)}{4\pi}(\sigma \Gamma + 3Hq\sigma+\dot{\sigma})N_{h},\label{eq:explicit}
\end{equation}
where we used our previous result $3Hq=\dot{V}_h/V_h - 3H$ together with Eqs. (\ref{eq:total}) and (\ref{eq:temp}). For $q=-1$ the horizon contribution vanishes, implying that the GSL is satisfied only if $\Gamma \geq 3H$ (assuming $\dot{\sigma}=0$). Therefore, as already noted in Eq. (\ref{eq:decelcond}), the particle production rate must be comparable to or greater than the Hubble expansion rate.

We would like to comment that in a matter creation scenario, the Friedmann constraint $3H^{2}=8\pi \rho$ and Eq. (\ref{eq:rho_creation}), lead to 
\begin{equation}
    \frac{\ddot{a}}{a}=-\frac{4\pi}{3}\left[\rho+3(p+p_{c})\right].\label{eq:accel}
\end{equation}
By combining Eq. (\ref{eq:pc_adiabatic}) with the condition $p=0$ in Eq. (\ref{eq:accel}), we find that $\ddot{a}>0$ is possible whenever $\Gamma > H$ \footnote{Alternatively, Eq. (\ref{eq:accel}) can be written as
\begin{equation}
    q = \frac{4\pi}{3H^{2}}\left[\rho+3(p+p_{c})\right] = \frac{1}{2}\left(1-\frac{\Gamma}{H}\right),\label{eq:q}
\end{equation}
where we have used $q= -\ddot{a}a/\dot{a}^2 = -\ddot{a}/(aH^{2})$ and Eq. (\ref{eq:pc_adiabatic}) together with $p=0$. Notice that condition $\Gamma > H$ leads to $q<0$.}. Hence, the requirements for thermodynamic consistency of the model are compatible with the dynamical behavior of cosmic evolution at late times.

To end this section, we focus on the adiabatic condition, $\dot{S}_{\mathrm{tot}}=0$. Taking into account $\dot{\sigma}=0$ in Eq. (\ref{eq:explicit}), we can straightforwardly solve for the particle production rate
\begin{equation}
   \Gamma = -\left(\frac{2\pi}{\sigma N_{h}H}(1+q)+3Hq \right),\label{eq:rate}
\end{equation}
where we can identify the horizon flux correction in the second term representing an extra contribution to the result found in \cite{cardenasphantom}; the sum of both terms must be negative to maintain $\Gamma >0$. In this case, the adiabatic condition $\dot{\sigma}=0$ enables the use of (\ref{eq:pc_adiabatic}). Thus, by choosing the case $p=0$, the creation pressure can be written as
\begin{equation}
    p_{c}=\left(\frac{1}{4\sigma N_{h}\rho}(1+q)+q \right)\rho,
\end{equation}
being similar to a barotropic equation of state featuring a dynamical state parameter, $p=w(t)\rho$. Since the quantities $\sigma$, $N_{h}$ and $\rho$ are strictly positive, the dimensionless phenomenological coupling parameter (which we formally define below as $\Omega_h \equiv (4\sigma N_{h}\rho)^{-1}$) satisfies $\Omega_h > 0$. Furthermore, to ensure that the creation pressure $p_c$ can successfully drive an accelerated expansion phase ($q < 0$) without triggering physical instabilities or violating the dominant energy condition, we impose the phenomenological upper bound $\Omega_h < 1$. This physical requirement ensures that the matter entropy remains bounded by its coupled energy density. Under these physically viable conditions, $0 < (4\sigma N_{h}\rho)^{-1} < 1$, and the condition $\left|q\right|>(4\sigma N_{h}\rho)^{-1}\left|1+q\right|$ could be fulfilled. For an expanding universe described by $q<0$ and $w=0$ together with the aforementioned condition for the absolute value of the deceleration parameter, we have a negative creation pressure, as expected. Following a well-known procedure, Eq. (\ref{eq:rho_creation}) can be recast in the standard form $\dot{\rho}+3H\rho(1+w_{\mathrm{eff}})=0$, where $w_{\mathrm{eff}}$ denotes the effective parameter state, defined as $w_{\mathrm{eff}} \equiv (p+p_{c})/\rho$. Using the corrected form of the creation pressure, we obtain the following expression for $w_{\mathrm{eff}}$ under the assumption $p=0$
\begin{equation}
    w_{\mathrm{eff}} = -\left[\frac{1}{4\sigma N_{h}\rho}(\left|q\right|-1)+\left|q\right|  \right],\label{eq:eff4}
\end{equation}
which, depending on the specific value of the deceleration parameter, may allow for a phantom–quintessence phase with $w=0$. Although the horizon flux correction broadens the scope of the results reported in \cite{cardenasphantom}, it ultimately yields the same scenarios for the late time cosmological evolution of an expanding universe.

\section{Final Remarks}
\label{sec:final}
In this work, we have revisited the derivation of the Friedmann equations from a thermodynamic perspective, applying the unified first law of thermodynamics to the apparent horizon of a flat FLRW universe. We extended the standard entropic cosmology framework to treat the region enclosed by the apparent horizon as an open thermodynamic system undergoing gravitationally induced particle creation.

The cornerstone of our formulation is the recognition that the apparent horizon volume is not comoving. By rigorously accounting for the particle exchange across this moving boundary, we demonstrated that the evolution of the particle number, and consequently the matter entropy, decomposes into two physically distinct contributions: genuine bulk particle production ($\Gamma N_h$) and a geometric horizon flux ($3HqN_h$) induced by the dynamics of the horizon itself. This decomposition allowed us to formulate a GSL that transparently balances horizon entropy, bulk creation, and boundary fluxes.

Therefore, as discussed in our work, the so‑called ``horizon flux'' is not merely a geometric artifact but a key physical element that restricts the particle production rate in a thermodynamically consistent way, driving an accelerated phase in cosmic evolution without the need to invoke dark energy.

Finally, it is worth noting that, in its present formulation, $w_{\mathrm{eff}}$ depends on the total particle number inside the apparent horizon, $N_h$, and on the specific entropy, $\sigma$, neither of which are directly observable cosmological quantities. We can map this thermodynamic framework to the kinematic redshift domain by using the standard relation for the deceleration parameter as a function of redshift $z$:
\begin{equation}
q(z) = -1 + \frac{1+z}{H(z)} \frac{dH}{dz} = -1 + \frac{1}{2}(1+z)\frac{d \ln H^2}{dz},
\end{equation}
where we have used the usual transformation, $1+z=a^{-1}(t)$. If we use our modified acceleration equation $\dot{H} = -4\pi\rho(1 + w_{\text{eff}})$ and employ the Friedmann constraint $\rho = \frac{3H^2}{8\pi}$, we can write $q=(1/2)(1+3w_{\text{eff}})$. Additionally, we can also obtain a differential equation governing the expansion history, given as
\begin{equation}
\frac{dH^2}{dz} = 3H^2 \frac{1 + w_{\text{eff}}(z)}{1+z}.
\end{equation}
To carry out parameter estimation using this model, the microscopic thermodynamic ratio present in Eq. (\ref{eq:eff4}) can be isolated as a dimensionless phenomenological coupling parameter, defined here as $\Omega_h(z) \equiv (4\sigma N_h \rho)^{-1}$. This parameter characterizes the ratio of the horizon flux correction to the total energy density. By rewriting Eq. (\ref{eq:eff4}) in terms of observables appropriate for the late-time accelerated phase ($q < 0$), we arrive at
\begin{equation}
 w_{\mathrm{eff}}(z) = \Omega_h(z) \left( 1 + q(z) \right) + q(z).
\end{equation}
Inserting this expression in $q(z)=(1/2)(1+3w_{\text{eff}}(z))$, we arrive to the condition $-q(z)(1+3\Omega_{h}(z))=1+3\Omega_{h}(z)$. This equality is only satisfied for $q=-1$ or $\Omega_{h}(z)=-1/3=\text{constant}$, which is unphysical due to our definition for the parameter $\Omega_{h}(z)$. Therefore, the adiabatic condition, $\dot{S}_{\mathrm{tot}}=0$, imposes $q=-1$ making the cosmic evolution of this framework indistinguishable from the standard cosmological model at background level. As discussed in \cite{Gohar_2026, Bhattacharjee_2026, Halder_2025, Gohar_2021}, a specific form of the production rate, $\Gamma$, is necessary to implement statistical analysis. In our framework, we must impose $\dot{S}_{\mathrm{tot}}>0$ to generate departures from the concordance model, which in turn requires an appropriate additional parametrization for $\Omega_{h}(z)$ and the deceleration parameter $q(z)$. Observe that, from Eq. (\ref{eq:rate}), it follows that $\Gamma = \Gamma(H,\dot{H})$, which constitutes an extension of the usual matter creation framework where one typically assumes $\Gamma = \Gamma(H)$. A detailed analysis of these issues is left for future work.

\begin{acknowledgments}
VHC acknowledges CEFITEV-UV for partial support. MC work was partially supported by S.N.I.I. (SECIHTI-M\'exico). SL acknowledges the FONDECYT grant N°1250969, Chile. 
\end{acknowledgments}

\onecolumngrid
\begin{center}
\begin{figure}[htbp!]
    \includegraphics[scale=0.45]{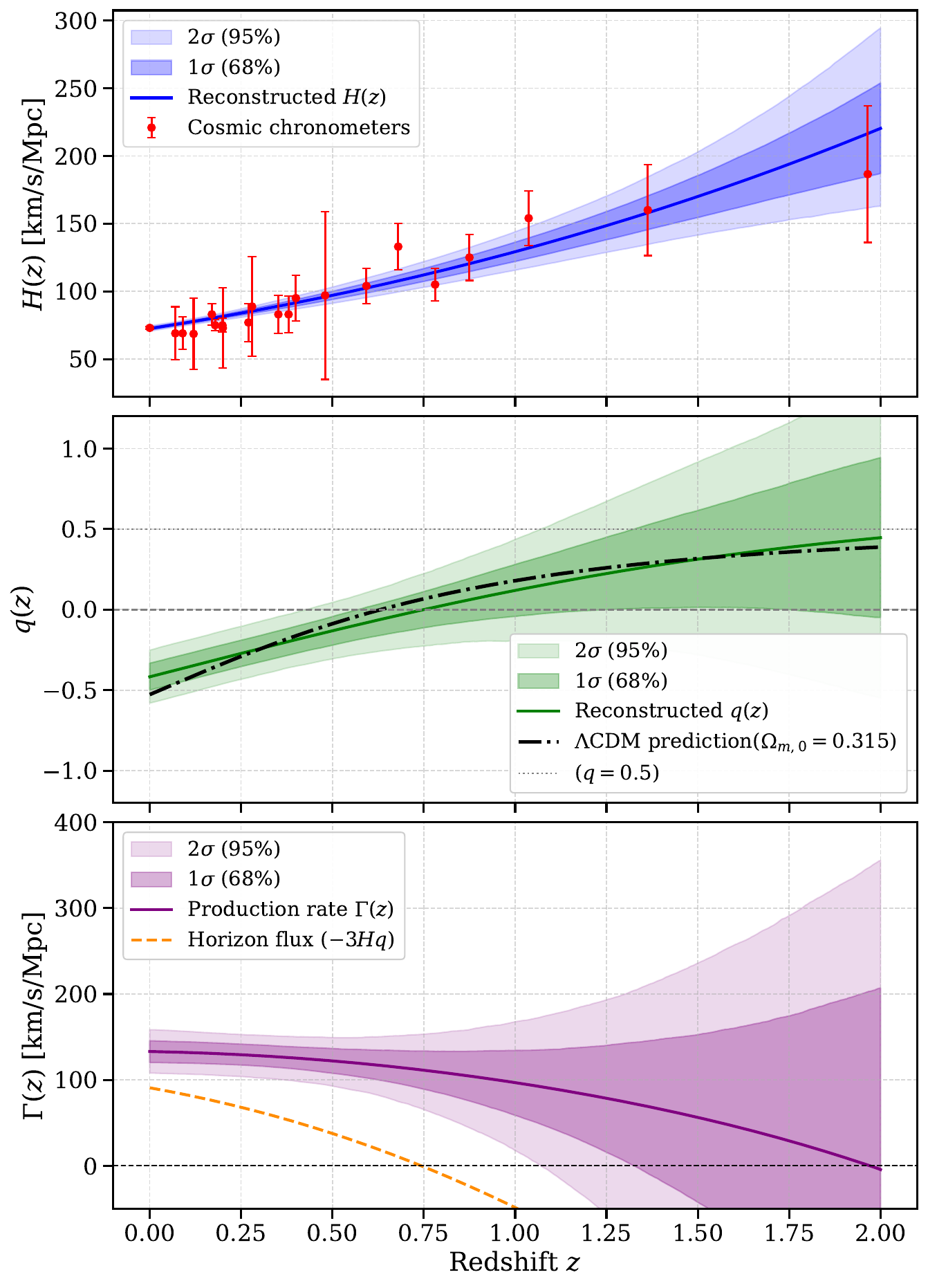}
    \caption{Reconstructed cosmological quantities in function of the redshift using the Cosmic Chronometers dataset. The upper panel corresponds to the Hubble parameter. The middle panel shows the deceleration parameter behavior and compares it against the $\Lambda$CDM prediction, we have considered the central value reported by the Planck collaboration for $\Omega_{m,0} = 0.315$ \cite{planck}. The lower panel corresponds to the production rate $\Gamma(z)$ and its consistency check against the horizon flux $-3Hq$ (yellow dotted line).}
    \label{fig:rec3}
\end{figure}
\end{center}
\twocolumngrid

\appendix
\section{Observational reconstruction of the model}
\label{sec:app}
To evaluate the viability of the thermodynamic framework developed in this work, we implement a non-parametric reconstruction that is independent of the background cosmological model. The objective is to isolate the kinematic evolution of the universe from observational data and map it onto the thermodynamic bounds required by our formalism, specifically the particle creation rate $\Gamma(z)$ and the entropy production. We employ the Gaussian Processes (GP) technique, see for instance Ref. \cite{Seikel_2012}, applied to the Cosmic Chronometers dataset \cite{di2023hubble}, which provides direct measurements of the expansion rate $H(z)$ that are independent of the cosmological model.
Assuming that the natural logarithm of the expansion rate follows a multivariate normal distribution, $\ln H(z) \sim \mathcal{GP}(0, K(z, z'))$, we optimize a Mat\'ern-type covariance kernel ($\nu = 5/2$) to reconstruct the continuous function $H(z)$. The main advantage of this formalism is the analytical computation of the derivative $dH/dz$, which allows us to calculate the kinematic deceleration parameter without presupposing any parametrization for dark energy:
\begin{equation}
    q(z) = -1 + \frac{1+z}{H(z)}\frac{dH}{dz}.
\end{equation}
The $1\sigma$ and $2\sigma$ statistical confidence regions were obtained through error propagation using $5000$ Monte Carlo realizations drawn from the GP posterior distribution.

According to our adiabatic creation model ($\dot{\sigma} = 0$) for a pressureless fluid ($p=0$), the background kinematics is intimately related to the microscopic dynamics through the modified acceleration equation. The production rate is written as follows from Eqs. (\ref{eq:pc}) and (\ref{eq:pc_adiabatic}) together with the First Friedmann equation (see also Eq. (\ref{eq:q}))
\begin{equation}
    \Gamma(z) = H(z) \left[ 1 - 2q(z) \right].\label{eq:ratea}
\end{equation}
As established previously, a consistency condition in an accelerating universe ($q < 0$) requires $\Gamma \geq -3Hq$. 

Our observational reconstruction, see Fig. (\ref{fig:rec3}), confirms that during the late-time acceleration regime ($z \lesssim 0.6$), the creation rate $\Gamma(z)$ reconstructed from observational data consistently remains above the geometric horizon flux limit ($-3Hq$). This result independently validates that the gravitationally induced particle creation mechanism can sustain the dark sector behavior without violating cosmic entropy.

\begin{center}
\begin{figure}[htbp!]
    \centering
    \includegraphics[scale=0.45]{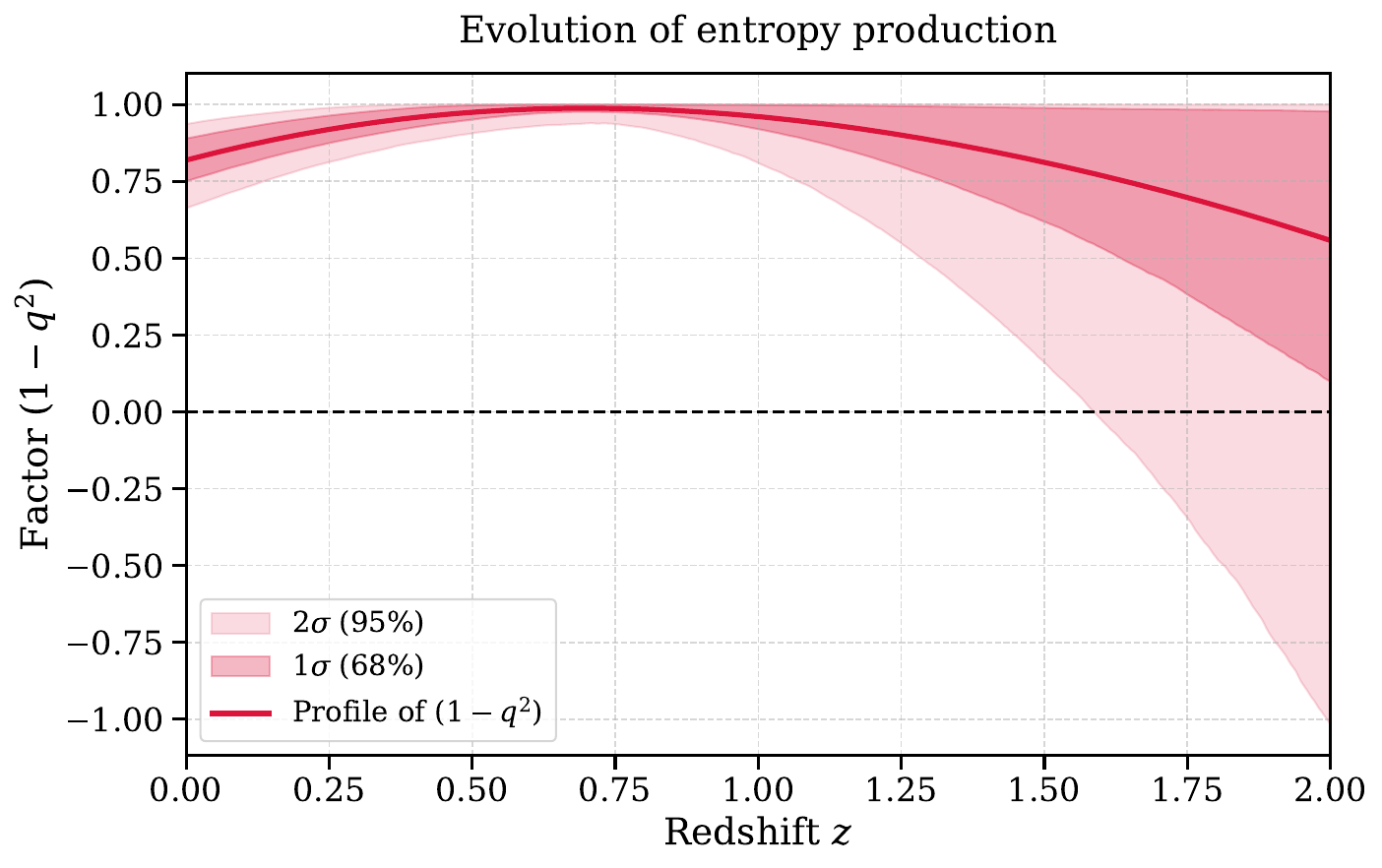}
    \caption{Behavior of kinematic factor $1-q^{2}$ along cosmic evolution.}
    \label{fig:entro}
\end{figure}
\end{center}

To further explore the thermodynamic robustness of the model, we derive an analytical expression for the total entropy production rate. If we insert the $\Gamma(z)$ function from Eq. (\ref{eq:ratea}) into the expression (\ref{eq:explicit}), we demonstrate that the total entropy depends on $\Omega_h \equiv (4\sigma N_h \rho)^{-1}$ as follows
\begin{equation}
    \dot{S}_{\text{tot}} =\frac{1}{2}(1-q^2)\left[ 1 + \frac{1}{3\Omega_h} \right],
\end{equation}
where we have considered $\dot{\sigma}=0$. Given that the underlying physical quantities demand $\Omega_h > 0$, the strict positivity of the entropy production is exclusively dominated by the kinematic factor $(1-q^2)$. As can be seen in Fig. (\ref{fig:entro}), the GP reconstruction of this factor reveals that the universe maximized its entropy production at the precise moment of the cosmic transition ($q=0$, at $z \simeq 0.6$). For late times, as the deceleration parameter asymptotically approaches $q \to -1$, the factor $(1-q^2)$ decreases towards zero, which is in agreement with the adiabatic scenario $\dot{S}_{\textrm{tot}}=0$ discussed in the work. Physically, this indicates that the entropy production gradually ceases, naturally and consistently driving the model towards the adiabatic limit of the de Sitter spacetime, without the risk of crossing into a non-physical regime ($\dot{S}_{\text{tot}} < 0$). 
\onecolumngrid
\twocolumngrid
This reconstruction supports the theoretical robustness of the proposed framework against observational evidence.

\bibliography{refs}

\end{document}